\documentclass{article}
\usepackage{spconf,amsmath,graphicx}
\usepackage{cite}

\usepackage{hyperref}

\title{A Deep Neural Network for Predicting Continuous Human EEG Across the Auditory Pathway in Response to Sound}
\name{Thomas J Stoll, Ross K Maddox\thanks{This work was supported in part by the
U.S. National Science Foundation under 2448814, the U.S. National
Institute of Health under T32 DC000011, and a gift from an anonymous donor.}}
\address{University of Michigan\\
Kresge Hearing Research Institute, Department of Otolaryngology--Head and Neck Surgery\\
Ann Arbor, MI 48109}
\begin{document}
%
\maketitle
\begin{abstract}
Computational models of auditory physiology commonly target specific responses or stages of the auditory pathway, limiting their ability to integrate findings across experimental paradigms and neural timescales. We present a foundation model of human auditory electrophysiology: a causal neural network trained to map binaural acoustic waveforms directly to high-sample-rate EEG. The model was trained on approximately 250 hours of EEG data from 92 subjects, with varied electrode montages and stimuli spanning tonebursts, speech, and music. We tested whether the model recovered effects of stimulus rate, frequency, and presentation method on auditory brainstem responses (ABRs); subcortical and cortical temporal response functions (TRFs) to continuous speech; and the click-evoked binaural interaction component (BIC). Predicted ABRs and TRFs reproduced established response morphology and stimulus-dependent effects, with model–grand-average correlations falling within the corresponding subject-level human distributions. The model-predicted BIC metrics closely resembled the values reported in the literature. These findings demonstrate that a single audio-to-EEG model can capture auditory physiology across paradigms and timescales, supporting future \textit{in silico} experimentation and hearing technology applications.
\end{abstract}
\begin{keywords}
Artificial neural networks, Auditory system, Deep learning, Electroencephalography
\end{keywords}
\section{Introduction}
\label{sec:intro}

Auditory neuroscience has traditionally relied on controlled experiments targeting specific mechanisms, subject populations, and stages of the auditory pathway. While this approach has elucidated many important aspects of auditory system function, small differences between studies make it difficult to synthesize findings across the literature into a unified understanding of the auditory brain\cite{dascoli_foundation_2026,yamins_using_2016}. Similarly, computational models have provided key insights towards a better understanding of the auditory system\cite{hrncirik_models_2023,johannesen_modeling_2022,kulasingham_predictors_2024,meddis_computer_2013,osses_vecchi_comparative_2022,shan_subcortical_2024,vasilkov_enhancing_2021,stoll_auditory_2025}, but conventional, hand-crafted models are impractical to extend across multiple response timescales and neural generators as complexity grows\cite{meddis_computational_2010,drakopoulos_convolutional_2021}. Recently, data-driven models have demonstrated promising results in auditory neuroscience\cite{dascoli_foundation_2026,yamins_using_2016,drakopoulos_convolutional_2021,drakopoulos_modelling_2025,sabesan_large-scale_2023,wingert_convolutional_2024,baby_convolutional_2021} and hearing technology, such as improved hearing aid algorithms\cite{wouters_evaluation_2026,wen_dconnear_2025,rosen_differentiable_2026,drakopoulos_optimal_2026}. These models provide a path towards a unified model for auditory neuroscience\cite{dascoli_foundation_2026,richards_deep_2019}, but existing models target one particular stage of the auditory pathway\cite{drakopoulos_modelling_2025,drakopoulos_convolutional_2021}, are built on invasive animal recordings\cite{drakopoulos_modelling_2025}, or model human performance in perceptual tasks rather than brain responses\cite{rosen_differentiable_2026}, potentially limiting their application in real-world scenarios. 

Here, we take a foundation-model approach to human auditory electrophysiology: we present a temporally causal encoder–decoder trained to transform raw acoustic waveforms to a continuous EEG response. The EEG datasets used here have a high sampling rate that enables the model to capture responses spanning early subcortical and later cortical timescales. The model is trained on many hours of EEG data from subjects that listened to stimuli including speech, music, and transient clinical stimuli. The heterogeneity of the training data provides a basis for the model to implicitly learn underlying physiological phenomena. We show that the model recapitulates established auditory phenomena across three experimental paradigms examining stimulus-dependent response characteristics, responses to naturalistic speech, and binaural integration.

\section{Methods}
\label{sec:methods}

\subsection{Datasets}

We curated 250 hours of high-sample-rate EEG data\cite{shan_subcortical_2024,stoll_auditory_2025,polonenko_parallel_2019,polonenko_exposing_2021} from 92 subjects (age 22.6$\pm$4.1 years, range 18–-38; 33 male, 59 female), spanning frequency-specific tonebursts, speech, music, and montages ranging from two to 64+ channels. All recordings were acquired at a $\ge$10 kHz sampling rate and were corrected for clock drift between the sound card and EEG systems. EEG was downsampled to 5 kHz, and causally high-pass filtered at 0.1 Hz (first-order Butterworth); audio was temporally aligned with the EEG, converted to Pascals and resampled to 40 kHz.

\subsection{Model Design}

Two-channel (stereo) audio was passed through a causal 24-band-per-ear filterbank spanning 100 Hz–16 kHz, log-compressed as $\log\left(|\mathbf{X}| + 1\right)$, and downsampled from 40 to 5 kHz using an average pooling layer. A nine-layer causal WaveNet\cite{oord_wavenet_2016} encoder used kernel size 8, dilations $2^d$ (where $d$ is the layer number, $d=0,\ldots,8$), and gated activations, with residual connections between each layer. A four-layer 1$\times$1-convolution bottleneck projected the representation onto 16 latent neural components, which were mapped to EEG using montage-specific spatial weights to yield the final neural prediction. The model was additionally conditioned on subject identity and demographic and audiometric metadata, although these effects are not analyzed here.

\subsection{Artifact Path}

A parallel linear pathway was used to model stimulus-locked electromagnetic artifact (a common issue in high-sampling-rate EEG recordings) during training using a 4 ms convolution over raw audio followed by subject- and recording-specific spatial projection. Symmetric padding provided the lookahead required to account for acoustic-tubing delay while the neural prediction pathway remained strictly causal. Neural and artifact predictions were summed during training, whereas the artifact pathway was disabled for all evaluations reported here.

\subsection{Training \& Optimization}

The network is optimized end-to-end and trained on recorded segments approximately one minute long. Training with MSE loss provided poor results. Instead, we computed the prediction loss in the short-time Fourier transform (STFT) domain to emphasize errors at higher frequencies so that the loss was not dominated by the less informative higher-power low frequency content in EEG. Let $y$ and $\hat{y}$ denote the target and predicted EEG signals, respectively. The STFT of the prediction error is defined as $S(f, \tau) = \text{STFT}(y - \hat{y})(f, \tau)$, where $f$ and $\tau$ denote the frequency bins and time frames, respectively. Importantly, the prediction error is calculated prior to taking the STFT, ensuring that phase (i.e., timing) differences are penalized in addition to amplitude differences. The log STFT prediction loss is then calculated as:
\begin{equation*}
L_{\text{STFT}} = \frac{1}{F} \sum_{f=0}^{F-1} 10 \log_{10} \left( \epsilon + \sum_{\tau} |S(f, \tau)|^2 \right)
\end{equation*}
where $F=251$ represents the total number of frequency bins and $\epsilon = 10^{-6}$ is a constant added for numerical stability. Optimization is performed using the Adam optimizer with a learning rate of $10^{-4}$ and batch size of 4. Gradient scaling (Automatic Mixed Precision) and gradient norm clipping (maximum norm of $1.0$) are used to stabilize training. Subject dropout was employed with a probability of $p=0.1$ to avoid over-reliance on subject embeddings, resulting in a ``default'' subject. Recent work has shown this default subject can be used for zero-shot predictions of population-level responses\cite{dascoli_foundation_2026}. Ten percent of data were held out for testing, and the model at the epoch with the lowest test loss was chosen for evaluation.

\begin{figure*}[!t]
    \centering
    \includegraphics[keepaspectratio]{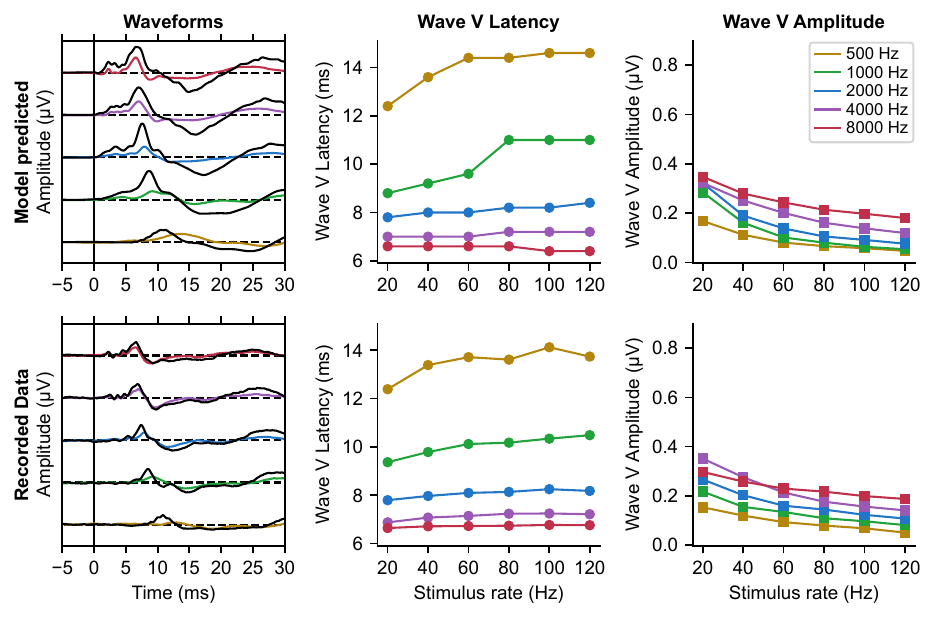}
    \caption{Predicted (top) and recorded\cite{polonenko_parallel_2019,polonenko_optimizing_2022} (bottom) pABR waveforms, wave V latencies, and amplitudes. Color denotes frequency; black denotes serial presentation.}
    \label{fig:pabr}
\end{figure*}

\subsection{Evaluation}

We evaluated the trained model with fixed weights and no fine tuning, using the default subject and neural pathway only. Evaluation waveforms were newly generated or held out from training, although their broader stimulus classes were represented during training. \textbf{pABR}: The parallel auditory brainstem response (pABR) utilizes randomized stimulus timing to collect responses to multiple frequencies in both ears at the same time\cite{polonenko_parallel_2019,polonenko_optimizing_2022}. Following this paradigm, we presented randomly-timed five-cycle tonebursts at 0.5, 1, 2, 4, and 8 kHz in parallel at 20--120 stimuli/s and responses were calculated through cross-correlation of the timing sequences (equivalent to averaging\cite{polonenko_parallel_2019}). We assessed frequency- and rate-dependent ABR wave V amplitude and latency and compared parallel with serial presentation (parallel: all frequencies presented at the same time; serial: one frequency presented at a time). Although the training data included pABR stimuli, newly generated random sequences were used here, and no training data included serial presentation. \textbf{Speech TRFs}: New speech from male and female talkers with the fundamental frequencies synthetically altered to be low or high\cite{polonenko_fundamental_2024} was presented to the model. Subcortical temporal response functions (TRFs) were estimated by filtering predictions from 30--2000 Hz and performing frequency-domain deconvolution using the glottal-pulse regressor\cite{polonenko_exposing_2021}. Cortical TRFs were estimated with ridge regression using the acoustic-envelope as the regressor after filtering predictions from 1--15 Hz\cite{bialas_mtrfpy_2023}. TRFs were estimated over lags of $-200$ to $500$ ms, with the ridge parameter of $\lambda=10$ selected via cross-validation on the recorded subjects' data. \textbf{ABR-BIC}: Periodic clicks were presented at 15.1 stimuli/s and 65 dB nHL to the left (L), right (R), or both (B) ears; the binaural interaction component (ABR-BIC)\cite{van_yper_binaural_2015} was calculated as $\text{ABR-BIC} = (\text{L} + \text{R}) - \text{B}$. While training data included both diotic and dichotic stimuli, it did not include clicks, monaural, or periodic stimuli.

\subsection{Performance Metrics}

To assess whether model predictions were representative of human responses, the predicted default subject responses were compared with human grand averages using Pearson correlation over the response windows (0--20 ms for pABRs and subcortical TRFs and 50--400 ms for cortical TRFs). Human baselines were formed by correlating each subject with a leave-one-out grand average. Model correlations were percentile-ranked within these distributions and compared using Crawford–Howell tests on Fisher-z-transformed correlations. Because recorded BIC waveforms were unavailable, predicted latency and amplitude were compared with published human summary statistics using Crawford–Howell single-case tests.

\section{Results}
\label{sec:results}
\begin{table}[!b]
\centering
\caption{Comparison of population-level model predictions with subject-level human correlation distributions..}
\label{tab:correlation_results}
\small
\begin{tabular}{lcccc}
\hline
\textbf{Condition} & $r_{\text{model}}$ & $p_{\text{CH}}$ & \%-tile & $r_{\text{human}} \pm \text{SD}$ \\
\hline
\textbf{pABR} & & & & \\
~~500 Hz & 0.604 & 0.352 & 20.7\% & $0.77 \pm 0.15$ \\
~~1000 Hz & 0.875 & 0.936 & 51.7\% & $0.84 \pm 0.11$ \\
~~2000 Hz & 0.944 & 0.154 & 96.6\% & $0.84 \pm 0.08$ \\
~~4000 Hz & 0.901 & 0.704 & 72.4\% & $0.85 \pm 0.09$ \\
~~8000 Hz & 0.931 & 0.198 & 89.7\% & $0.83 \pm 0.10$ \\
\hline
\textbf{Subcortical TRF} & & & & \\
~~Male--Low F0 & 0.870 & 0.256 & 93.3\% & $0.72 \pm 0.15$ \\
~~Male--High F0 & 0.438 & 0.935 & 46.7\% & $0.39 \pm 0.23$ \\
~~Female--Low F0 & 0.840 & 0.309 & 80.0\% & $0.52 \pm 0.32$ \\
~~Female--High F0 & 0.447 & 0.905 & 46.7\% & $0.45 \pm 0.21$ \\
\hline
\textbf{Cortical TRF} & & & & \\
~~Male--Low F0 & 0.895 & 0.165 & 93.3\% & $0.67 \pm 0.17$ \\
~~Male--High F0 & 0.932 & 0.019 & 100.0\% & $0.63 \pm 0.19$ \\
~~Female--Low F0 & 0.791 & 0.816 & 73.3\% & $0.69 \pm 0.20$ \\
~~Female--High F0 & 0.845 & 0.165 & 93.3\% & $0.63 \pm 0.18$ \\
\hline
\end{tabular}
\end{table}

\begin{figure}[htb]
    \centering
    \includegraphics[width=\linewidth,keepaspectratio]{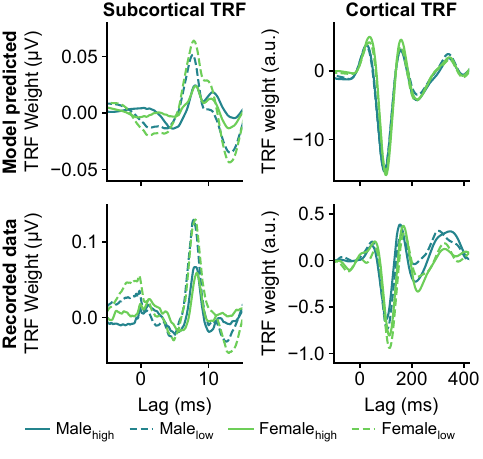}
    \caption{Predicted (top) and recorded\cite{polonenko_fundamental_2024} (bottom) subcortical and cortical TRFs.}
    \label{fig:trf}
\end{figure}

Predicted pABR and speech TRF waveforms closely resembled the corresponding human grand averages (Figs.~\ref{fig:pabr}--\ref{fig:trf}). The model reproduced frequency-dependent pABR morphology, rate-dependent wave V amplitude and latency\cite{polonenko_optimizing_2022}, serial--parallel differences\cite{polonenko_parallel_2019}, and the effect of larger subcortical TRFs to low-F0 speech for both talkers\cite{polonenko_fundamental_2024}.

\begin{figure}[htb]
    \centering
    \includegraphics[width=\linewidth,keepaspectratio]{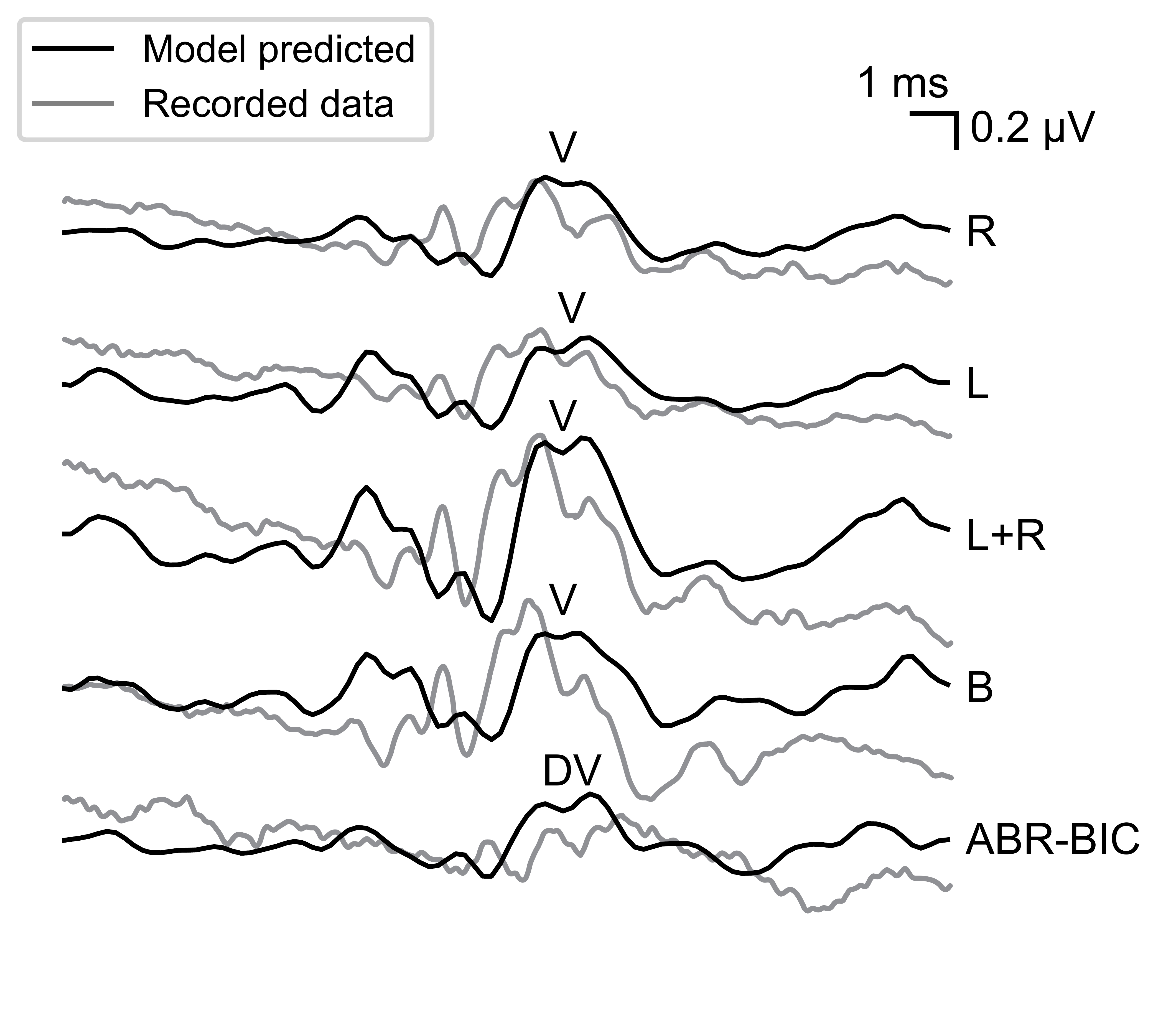}
    \caption{Predicted monaural (L, R), binaural (B), summed-monaural ($\text{L}+\text{R}$), and binaural interaction component [$\text{ABR-BIC}=(\text{L}+\text{R})-\text{B}$] click responses. An example individual's response\cite{van_yper_binaural_2015} is shown in gray, model predicted responses are overlaid in black.}
    \label{fig:bic}
\end{figure}

Model-human correlations did not differ significantly from the subject-level human distribution in all pABR conditions and all but one TRF condition  Table~\ref{tab:correlation_results}). The exception was the male high-F0 cortical TRF, for which the model prediction correlated more strongly with the grand average than did individual subjects ($p_{\mathrm{CH}}=0.019$). Model rate–amplitude curves closely followed the human group trends at every frequency ($r\geq0.961$, $p_{\mathrm{CH}}\geq0.05$) except for the 8-kHz curve, where model correlation again exceeded the subject-level distribution ($p_{\mathrm{CH}}=0.048$). Rate–latency curve correlations did not differ significantly from their human distributions. Notably, the model correlation was higher than the human-to-human mean across all conditions except for the 500 Hz pABR waveforms and the Female High subcortical TRF.

The model also replicated known binaural integration effects, showing ABR-BIC morphology similar to an example response from the literature\cite{van_yper_binaural_2015} (Fig.~\ref{fig:bic}). Its latency (6.40 ms) was within the published range (5.58--6.90 ms), whereas its amplitude (0.439~\textmu V) exceeded the reported range (0.13--0.36~\textmu V). Still, neither value differed significantly from the published human sample under Crawford--Howell single-case comparisons (latency: $t_{\mathrm{CH}}(16)=0.933$, $p=0.364$; amplitude: $t_{\mathrm{CH}}(16)=1.574$, $p=0.135$).

Some differences between model predictions and responses from recorded data are seen in the size of model-predicted responses from serial presentation that are larger than recorded responses (Fig. \ref{fig:pabr}) and model-predicted TRFs that are smaller than recorded (Fig. \ref{fig:trf}). Both of these differences may be due to the model overestimating effects of neural adaptation or efferent activation. Differences in BIC waveforms are not directly interpretable, since we do not have access to grand average waveforms (only a single example subject from \cite{van_yper_binaural_2015} is shown in Fig. \ref{fig:bic}).

\section{Discussion}
\label{sec:discussion}

We demonstrated that a single audio-to-EEG model recovered established human auditory phenomena for both transient and continuous stimuli and across subcortical and cortical timescales. Without evaluation-specific fine-tuning or subject calibration, the model reproduced stimulus-driven effects in pABRs, subcortical and cortical speech TRFs, and the click-evoked BIC. Correlations between model predictions and recorded grand averages generally fell within—and occasionally above—the corresponding distributions of subject-level correlations. The success of this model is driven by the diversity of the stimuli and subjects represented in the training data, supporting a foundation-model approach to human auditory electrophysiology.

Future work will examine additional auditory phenomena, characterize and validate the subject-specific conditioning, and expand the training corpus—particularly across age and hearing ability. Whereas existing EEG foundation models have focused on extracting features from EEG\cite{kuruppu_eeg_2026}, the model presented here predicts auditory-evoked EEG from sounds. Thus, as the model matures, it may provide a practical platform for \textit{in silico} experimentation and for developing experimental paradigms and analysis methods before undertaking costly and time-consuming data collection. Additionally, because the neural pathway is temporally causal, the model could be integrated into streaming pipelines that evaluate or optimize hearing-aid processing using predicted neural responses or as a front-end for automated feature extraction methods.

\vfill\pagebreak

\begingroup
\small
\bibliographystyle{IEEEbib}
\bibliography{refs}
\endgroup

\end{document}